\documentclass[twocolumn,twocolappendix,tighten,times,astrosymb]{aastex631}

\usepackage{amsmath}
\usepackage{booktabs}
\usepackage{soul}
\usepackage{microtype}
\usepackage[english]{babel}
\usepackage{cancel}
\usepackage{soul}
\hypersetup{linkcolor=blue,citecolor=blue,filecolor=blue,urlcolor=blue}

\newcommand{\be}{\begin{equation}}
\newcommand{\ee}{\end{equation}}

\newcommand{\txs}{TXS~0506+056}
\long\def\exclude#1{}

\shorttitle{Coronal neutrinos from TXS 0506+056}
\shortauthors{}
\begin{document}

\title{\large Can the neutrinos from TXS 0506+056 have a coronal origin?}

\correspondingauthor{damianofg@gmail.com}
\author[0000-0003-4927-9850]{Damiano F. G. Fiorillo}
\affiliation{Deutsches Elektronen-Synchrotron DESY, Platanenallee 6, 15738 Zeuthen, Germany}

\author[0009-0000-9401-1971]{Federico Testagrossa}
\affiliation{Deutsches Elektronen-Synchrotron DESY, Platanenallee 6, 15738 Zeuthen, Germany}

\author[0000-0001-6640-0179]{Maria Petropoulou}
\affiliation{Department of Physics, National and Kapodistrian University of Athens, University Campus Zografos, GR 15784, Athens, Greece }
\affiliation{Institute of Accelerating Systems \& Applications, University Campus Zografos, Athens, Greece}

\author[0000-0001-7062-0289]{Walter Winter}
\affiliation{Deutsches Elektronen-Synchrotron DESY, Platanenallee 6, 15738 Zeuthen, Germany}

\begin{abstract}
The blazar TXS 0506+056 has been the first astrophysical source associated with high-energy astrophysical neutrinos, and it has emerged as the second-most-prominent hotspot in the neutrino sky over ten years of observations.  Although neutrino production in blazars has traditionally been attributed to processes in the powerful relativistic jet, the observation of a significant neutrino flux from NGC 1068 -- presumably coming from the Active Galactic Nucleus (AGN) corona -- suggests that neutrinos can also be produced in the cores of AGN. This raises the question whether neutrino production in TXS~0506+056 is also associated with the core region. We study this scenario, focusing on the hypothesis that this blazar is a masquerading BL Lac, a high-excitation quasar with hidden broad emission lines and a standard accretion disk.  We show that magnetic reconnection is an acceleration process necessary to reach tens of PeV proton energies, and we use observationally motivated estimates of the X-ray luminosity of the coronal region to predict the emission of secondaries and compare them to the observed multi-wavelength and neutrino spectra of the source. We find that the coronal neutrino emission from TXS 0506+056 is too low to describe the IceCube observed neutrinos from this AGN, which in turn suggests that the blazar jet remains the preferred location for neutrino production.
\end{abstract}


\section{Introduction}

The discovery of a diffuse astrophysical neutrino flux in the 10~TeV-10~PeV range  by IceCube marked a milestone in multi-messenger physics~\citep{IceCube_2013, Aartsen_2013}. Identifying the sources contributing to this flux remains a fundamental open question. The primary challenge lies in the fact that most detected neutrinos cannot be traced back to specific point sources with certainty, while a diverse range of astrophysical objects has the potential to serve as neutrino emitters. 

Searching for spatial and temporal correlations among the astrophysical neutrinos and catalogs of known sources, either steady or transient, has led to the identification of a few candidate sources. The first among them has been the IceCube detection in 2017 of an astrophysical neutrino with an energy of approximately 290 TeV in spatial and temporal coincidence with an electromagnetic flare from the blazar \txs~\citep{IC_TXS_2017}. Blazars are active galactic nuclei (AGN), namely galaxies exhibiting powerful multi-wavelength emission powered by accretion onto a central supermassive black hole \citep[see, e.g.,][]{Marconi_2004_SED,Ghisellini_2013,Padovani:2017zpf}, with a powerful jet pointing towards the Earth. At the time of neutrino detection, \txs\ was undergoing a six-month-long flare in its gamma-ray emission, coming from the jet region of the blazar. This coincident detection can be modeled assuming that neutrinos are produced in the blazar jet~\citep[e.g.][]{Ansoldi2018, Keivani:2018rnh,Liu2019,Gao2019,Cerruti2019}. A search through the archival IceCube data revealed a neutrino excess from the direction of the blazar in the 2014/2015 time frame, which, however, was not accompanied by an electromagnetic flare~\citep{IC_TXS_2014}. This makes an interpretation of the 2014/2015 flare quite challenging, since neutrino emission should be accompanied by a gamma-ray flux which, even if reprocessed to lower energies, should still show up in some part of the electromagnetic spectrum~\citep[e.g.][]{Murase2018, Reimer2019, Rodrigues2019, Petropoulou:2019zqp}. Finally, an analysis of 10-year IceCube observations of muon neutrinos has revealed that \txs\ is spatially coincident with the second brightest hotspot in the neutrino sky, suggesting that this blazar may also be a steady emitter of neutrinos up to PeV energies.
All this observational evidence has been almost always interpreted in terms of neutrino production in the powerful blazar jet.

Meanwhile, a different paradigm for neutrino emission from AGN has been suggested by the observation of a neutrino excess at energies of a few TeV from the direction of the Seyfert galaxy NGC 1068 in the 10-year IceCube measurements \citep{Aartsen_Icecube_2020,IceCube-NGC1068}. As this detection was not accompanied by corresponding TeV gamma-ray emission, with strong upper limits placed by the MAGIC telescope~\citep{MAGIC-UL-NGC1068}, it was suggestive of a gamma-ray opaque neutrino production region~\citep{Murase:2022dog} that is likely very close to the central black hole. A natural candidate emerges as the \textit{corona} of the AGN, responsible for the non-thermal X-rays observed from a vast majority of non-jetted AGN. While the shape and properties of the corona can vary widely across models~\citep[see e.g.][and references therein]{Cackett2021}, it usually has a typical size of a few up to several tens of Schwarzschild radii, and contains a dense X-ray field originating from Comptonization of optical/UV disk photons by energetic electrons~\citep[e.g.,][]{1991ApJ...380L..51H, Merloni_2001, Wilkins_2015}. The X-ray field can act both as an absorber of gamma-rays, providing the required gamma-ray opacity, and as a target for proton-photon collisions, in which neutrinos are produced~\citep[e.g.,][]{1981MNRAS.194....3B}. The corona is thus a natural neutrino production site provided that a mechanism to accelerate protons to very high energies is available. Various scenarios for proton energization have been proposed, including diffusive shock acceleration~\citep{Inoue_2020}, stochastic gyroresonant acceleration~\citep{Murase:2019vdl}, non-resonant acceleration in strongly magnetized turbulence~\citep{Fiorillo:2024akm,Lemoine:2024roa}, magnetized reconnection~\citep{Fiorillo:2023dts,Karavola:2024uui}, as well as a two-step scenario with pre-acceleration from reconnection and re-acceleration in turbulence~ \citep{Mbarek:2023yeq}.

These recent developments have demonstrated that non-jetted AGN may be efficient emitters of neutrinos. This naturally raises the question of the respective contributions of the corona and the jet to the neutrino signal expected from blazars. \txs, undoubtedly, stands out as an ideal source for addressing this question. \cite{Kimura:2020srn} concluded that the neutrino luminosity from a hot accretion flow in \txs \ should be smaller than about $10^{44}$~erg/s for a magnetically arrested disk (MAD) with sub-Eddington accretion. However, 
recently, \cite{2024arXiv241114598K, Yang:2024bsf} proposed that 
multi-messenger observations of \txs\ might be interpreted by neutrino production in the core region of the AGN, namely a compact region close to the black hole, such as the corona or the inner accretion flow, rather than the blazar jet as previously assumed. {\cite{Yang:2024bsf}, using a modeling similar to~\cite{Kimura:2020srn}, concludes that the 10-years-averaged neutrino signal from \txs might indeed be produced in the hot accretion flow, and that even the 2014/15 neutrino flare might be explained if temporary super-Eddington accretion is assumed. A core neutrino production} would have consequences not only on the specific question of the \txs\ neutrinos, but more broadly on the blazar contribution to the diffuse neutrino sky compared to non-jetted AGN. 

In this work, we re-examine the corona hypothesis for neutrino production in \txs. We use observational properties of the blazar to infer its accretion-driven X-ray luminosity. Within the context of the scenario of~\cite{Fiorillo:2023dts} (hereby denoted F24), in which non-thermal particles are energized within a magnetic reconnection layer close to the central black hole, we construct a model for coronal neutrino production consistent with the energetics and the X-ray luminosity in the corona. By comparing to the potential emission from a jet (blazar), we then critically assess the possibility that the observed neutrinos from \txs\ come from the corona.

\section{Coronal model of neutrino emission}\label{sec:coronal_model}

\begin{figure*}
    \includegraphics[width=\textwidth]{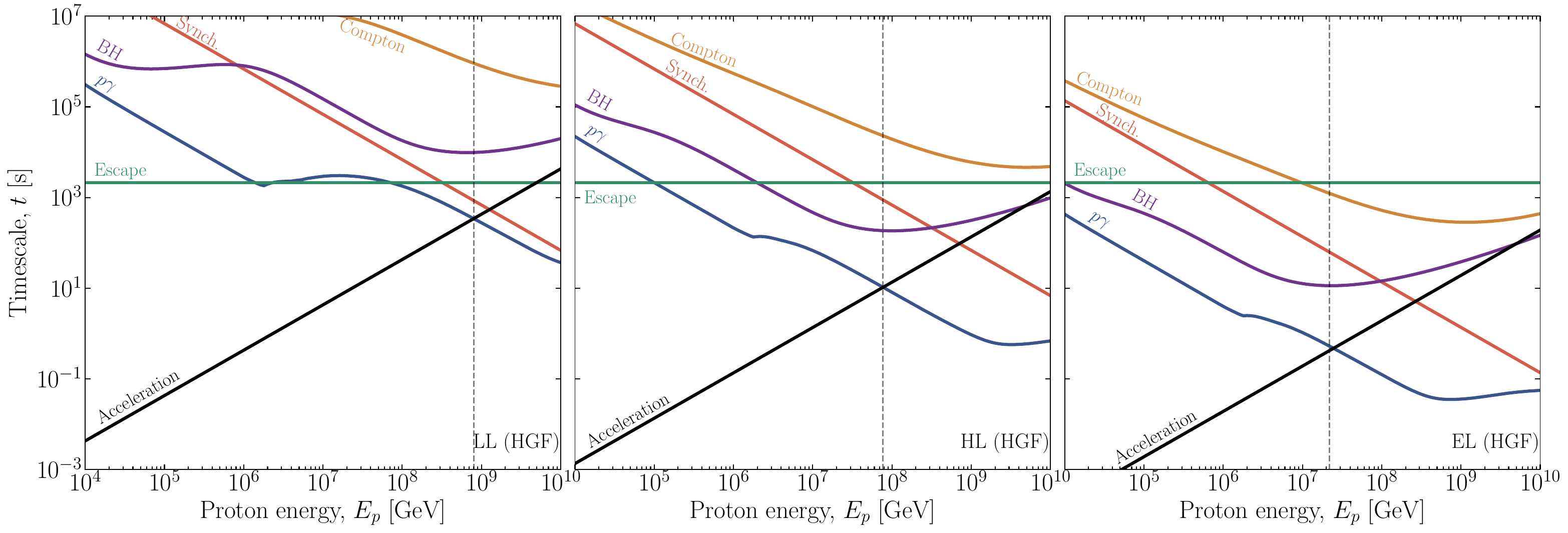}
    \caption{Timescales for acceleration, escape, and cooling processes of protons. We highlight by a vertical dashed line the value of $E_{p,\rm rad}$ at which the proton injection spectrum becomes exponentially suppressed due to cooling becoming more rapid than acceleration.}\label{fig:timescales}
\end{figure*}

We summarize the main properties of the coronal region and the mechanism by which non-thermal particles may be accelerated and interact within it. The corona is assumed to be a compact region located close to the central supermassive black hole, with dimensions up to a few tens of gravitational radii $r_g=G M/c^2$, where $G$ is gravitational constant, $c$ is the speed of light, and $M$ is the mass of the central supermassive black hole. 
For \txs, we use $M\simeq 3\times 10^8\;M_{\odot}$~\citep{Padovani:2019xcv}, so $r_g\simeq 4.4\times 10^{13}$~cm. 
The corona itself is permeated by a dense X-ray field, generated by Comptonization of low-energy photons -- either from the accretion disk or internal to the corona -- from energetic leptons~\citep{1979Natur.279..506S,1980A&A....86..121S,1991ApJ...380L..51H,1993ApJ...413..507H} or from the bulk motion of turbulent plasma~\citep{Groselj_2023arXiv} or bulk motion of plasmoids in reconnection layers~\citep{Beloborodov:2017njh,Sironi:2019sxv,Sridhar:2021bvf,Sridhar:2022ojr}. The X-ray field also plays a key role in non-thermal dynamics, since it acts as a target for $p\gamma$ and $\gamma\gamma$ interactions, rendering the corona opaque to high-energy gamma rays and an efficient converter of proton energy into neutrinos and electromagnetic radiation. The optical thickness of the corona to said processes depends crucially on the size of the corona and especially on the mechanism by which non-thermal particles are accelerated. 

For most of this work, we will consider the scenario proposed in F24, in which the corona is identified with a magnetized reconnection layer and protons are accelerated fast by the reconnection electric field. This scenario allows for neutrino production up to PeV energies, as required to explain the \txs\ neutrino observations. In Sec.~\ref{sec:discussion} we will also discuss how slower forms of acceleration, due to scattering by magnetized turbulence in the accretion flow, are expected to produce much less energetic neutrinos.

Within the corona, a population of leptons must be established with a moderate optical thickness $\tau_T\simeq 0.5$, as suggested by the shape of the X-ray spectrum produced after Comptonization~\citep{Wilkins:2014caa,Petrucci:2020cda,Tripathi:2022bog} as well as from radiative particle-in-cell (PIC) simulations of reconnection coupled with Monte Carlo radiative transfer calculations~\citep{Sridhar:2021bvf,Sridhar:2022ojr}. The number density of pairs in a corona with size $R$ can be then inferred by the typical Thomson optical depth, $n_e \simeq \tau_T/(\sigma_T R) \simeq 10^{10} \tau_{T, -0.3} \, R_{13.8}^{-1}~\rm cm^{-3}$, where we introduce the notation $q_x = q/10^x$ (in cgs units). On the other hand, the proton density $n_p$ can be much smaller than the electron one if the corona is pair-dominated, as expected in magnetospheric current sheets (F24). The properties of accelerated particles in reconnection layers depend on the magnetization defined separately for each species as  $\sigma_{e,p}=B^2/(4\pi n_{e,p} m_{e,p} c^2)$, where $B$ is the strength of the reconnecting magnetic field~\citep{werner_17,Zhang:2021akj,Zhang_2023arXiv,Chernoglazov_2023arXiv}. Therefore, if the corona is powered by magnetic energy dissipation in magnetospheric current sheets we expect $\sigma_p \gg \sigma_e > 1$.

In \cite{Zhang:2021akj,Zhang:2023lvw} a model of non-thermal particle acceleration in reconnection layers was proposed, which was based on findings of PIC simulations. A detailed investigation of the radiative signatures of accelerated leptons in this framework was recently presented in \cite{Stathopoulos:2024aro}. Here, we extend the findings of \cite{Zhang:2021akj,Zhang:2023lvw} to describe the acceleration of protons, as detailed in F24 for Seyfert galaxies. Similarly to leptons, protons gain energy in the upstream region during a phase of active acceleration
on a timescale $t_{\rm acc}\simeq E_p/(\beta_{\rm rec} eBc)$, where $\beta_{\rm rec}\simeq 0.1$ is the speed of plasma inflow within the layer in units of the speed of light. 
This estimated acceleration time is much closer to that assumed in~\cite{Yang:2024bsf}. Protons are captured by plasmoids of reconnected plasma (downstream region) on a comparable timescale, after which they are no longer actively accelerated. Non-thermal protons are therefore injected from the upstream region into the downstream region. For practical purposes, we consider only protons within the downstream region, where they spend most of the time, leaving the system (current sheet) after a much longer timescale $t_{\rm esc}\simeq L/c$, where $L$ is the size of the reconnection layer. 

We first discuss the properties of the proton injection spectrum. 
PIC simulations\footnote{These do not include cooling processes for relativistic protons.} show that the differential proton number density can be described by a broken power law, $dn_p/dE_p\propto E_p^{-1}$ for $E \le E_{p,\rm br}\simeq \sigma_p m_p c^2$ and $dn_p/dE_p\propto E_p^{-s}$ for $E > E_{p,\rm br}$. Protons accelerated by the reconnection process are in rough energy equipartition with the reconnecting magnetic field. The post-break power-law index $s$ ranges from $2$ to $3$ depending on the strength of the non-reconnecting component of the magnetic field, known as the guide field. 
A strong guide field ($\gtrsim 0.5 B$) renders acceleration less efficient and therefore produces a soft power law with $s\simeq 3$, while $s\simeq 2$ for weak guide fields. A strong guide field is favored from the IceCube detection of NGC~1068 neutrinos~\citep{Fiorillo:2023dts} and has been observed in the reconnection layers formed at the black hole jet sheath~\cite{Sridhar:2024rii}; for our purposes, however, the difference between the two cases is not essential, as we will discuss. At an energy large enough that the acceleration rate is comparable with radiative cooling rate, which we define as $E_{p,\rm rad}$, the spectrum exhibits a cutoff. This summarizes the properties of the proton injection spectrum in the downstream region. 

The overall scenario is qualitatively captured by the cooling and acceleration timescales in Fig.~\ref{fig:timescales}; here we consider three scenarios -- a low-luminosity (LL), high-luminosity (HL), and an extreme luminosity (EL) scenario specific for the 2014/15 neutrino flare of \txs -- whose detailed parameters are summarized in Sec.~\ref{sec:multimessenger}. The acceleration rate in the upstream region is the fastest process up to $E_{p,\rm rad}$ where photohadronic scattering rapidly cools protons. For all scenarios we find $E_{p, \rm rad} >10 \rm~PeV$. Once protons are injected in the downstream region, they are not further energized, and remain confined within the plasmoids in the layer. These plasmoids escape from the current sheet with relativistic speeds on a timescale $t_{\rm esc}$ equal to the light-crossing time of the layer.

High-energy neutrinos and gamma rays are injected into the corona through $p\gamma$ interactions. While neutrinos leave the region unimpeded, gamma rays are reprocessed by photon-photon pair production and subsequent radiative cooling of pairs, leading to a radiative cascade across the entire electromagnetic spectrum. The majority of the energy is finally released in the form of radiation below the MeV energy scale, as well as in the production of a large number of pairs that ultimately cool within the reconnection layer.

\section{Observational evidence for the coronal X-ray luminosity}\label{sec:observational_X_ray}
Direct observation of the coronal X-ray emission in blazars is challenging because of the highly beamed X-ray emission from the relativistic jet. In this section, we put together different pieces of information to infer an approximate value of the coronal luminosity in TXS~0506+056.

\txs\ is a masquerading BL~Lac, namely a high-excitation active galaxy whose thermal accretion-related emission is hidden by the jet's non-thermal emission~\citep{Giommi_2013, Padovani:2019xcv}. The bolometric accretion disk luminosity, which cannot be directly measured in these blazars, can be estimated using empirical relationships between the accretion-driven bolometric luminosity, $L_{\rm bol}$ and the luminosities of specific emission lines~\citep[see e.g.][and references therein]{Punsly_2011}. Applying a relation involving the luminosities of the $\rm O~II$ and $\rm O~III$ lines, \cite{Padovani:2019xcv} estimated the bolometric (thermal) luminosity to be $L_{\rm bol} \sim 1.7\cdot 10^{45}$~erg s$^{-1}$. Furthermore, \cite{Padovani:2019xcv} used an empirical relationship between the black hole mass and the absolute optical magnitude of the host galaxy's bulge \citep{Mclure_Dunlop_2002} to derive $M \simeq 3 \cdot 10^8 M_\odot$, which translates to an Eddington luminosity $L_{\rm Edd} \simeq 4 \cdot 10^{46}$~erg s$^{-1}$. Therefore, the accretion-driven luminosity of TXS~0506+056 has an Eddington ratio $L_{\rm bol}/L_{\rm Edd} \simeq 0.04$, suggestive of the presence of an optically thick and geometrically thin disk. Half of the bolometric luminosity may be taken as a proxy of the accretion disk luminosity, motivated by the spectral energy distributions (SEDs) of typical quasars~\citep{Padovani:2019xcv}. Alternatively, whenever narrow emission lines are available, the disk luminosity can be inferred assuming a typical covering fraction of the BLR (namely the ratio of BLR to disk luminosity) of 10\%. For TXS~0506+056 the logarithmic mean of the two methods yields $L_{\rm disk} \approx 3\cdot 10^{44}$~erg~s$^{-1}$~\citep{Padovani:2019xcv, Rodrigues_2024}. Finally, the high signal-to-noise ratio optical spectrum obtained by \cite{Paiano_2018} resembles that of a typical Seyfert 2 galaxy in terms of emission line ratios.

Quantifying the relation between the UV disk luminosity and the X-ray non-thermal emission from the corona in non-jetted AGN has been an active topic of research. Several empirical relations have been derived between the monochromatic UV and X-ray luminosities, or the bolometric luminosity and the 2-10 keV X-ray luminosity, or the optical-to-X-ray spectral index $\alpha_{\rm ox}$ and the Eddington ratio for bright AGN and changing-look quasars \citep[e.g.,][]{Lusso_2010, Lusso_2011, Arcodia_2019, Ruan_2019}.

We estimate the X-ray luminosity of the corona in \txs\ using two methods. First, we estimate the 3000~$\textup{~\AA}$ luminosity using Eq.~(10) from \cite{Runnoe_2012}, $L_{\rm UV} \approx 2.5 \cdot 10^{44}$~erg~s$^{-1}$. Then, using Eq.~(12) from \cite{Arcodia_2019} we estimate the 2~keV X-ray luminosity, $L_{2~\rm keV} \approx 6.4 \cdot 10^{43}$~erg~s$^{-1}$. Assuming that the corona spectrum is a power law, ${\rm d}n_X/{\rm d}E_X\propto E_X^{-2}$, we can derive the 2-10~keV X-ray luminosity, $L_{2-10~\rm keV} \approx 10^{44}$~erg~s$^{-1}$, and the broadband (0.1-100 keV) X-ray luminosity, $L_X \approx 4.4 \cdot 10^{44}$~erg~s$^{-1}$. Alternatively, we may use the empirical relation for the 2-10 keV X-ray luminosity obtained by \cite{Runnoe_2012} for a sample of radio-loud AGN (see Eq.~(14) therein), which yields $L_{2-10~\rm keV} \approx 8.2 \cdot 10^{42}$~erg~s$^{-1}$ and $L_{X} \approx 3.5 \cdot 10^{43}$~erg~s$^{-1}$. The estimates quoted above, which do not account for the statistical uncertainties of the empirical relations, differ approximately by a factor of 10.

Our inferred X-ray luminosity is much lower than the observed one from \txs. More specifically, \textit{Swift}-XRT observations prior and after the 2017 flare have shown that the 0.3-10~keV X-ray luminosity varies from approximately $3.8 \cdot 10^{44}$~erg~s$^{-1}$ to $3\cdot 10^{45}$~erg~s$^{-1}$, and the photon index of the spectrum varies from 1.3 to 2.7~\citep{Petropoulou_2020, Acciari_2022}. Furthermore, NuSTAR observations performed during 2018--2019 show that the hard (3-78 keV) X-ray luminosity is $\sim (1.5-3) \cdot 10^{45}$~erg~s~$^{-1}$. Although the most likely origin for the variable X-ray emission in TXS~0506+056 is the jet, recent studies have adopted the hard (15-55~keV) X-ray luminosity as a proxy for coronal emission \citep{2024arXiv241114598K, Yang:2024bsf}, thus overestimating its emission. Motivated by the discussion in this section, we conclude that the steady X-ray coronal luminosity ranges instead between $L_{X}=4\times 10^{43}$~erg/s and $L_X=4\times 10^{44}$~erg/s.

\section{Multi-messenger emission from TXS~0506+056}\label{sec:multimessenger}

In this section we obtain the neutrino and electromagnetic coronal emission assuming the reconnection scenario outlined in Sec.~\ref{sec:coronal_model}. Numerical calculations of the multi-messenger emission are performed with the code \texttt{AM$^3$}~\citep{Klinger:2023zzv}.

We consider the downstream region of the reconnection layer as the neutrino production site. The coronal X-ray field is modeled as a power law with photon index 2 between 0.1 and 100 keV and broadband X-ray luminosity $L_{X}$. A magnetospheric reconnection layer can be thought of as a thin rectangular sheet of typical dimension $L$ and thickness $\beta_{\rm rec} L$. Because the code is tailored for spherically symmetric emitting regions, we follow the discussion in Appendix~B of F24 and choose a radius $R$ corresponding to a sphere with the same volume as the slab.

A fraction $\eta_X=0.5 \, \eta_{X, -0.3}$ of the Poynting energy flux through the reconnection layer $c\beta_{\rm rec} B^2/4\pi$, is transformed into an X-ray energy flux escaping from the corona $L_{X}/4\pi R^2$. Then, the magnetic field strength can be expressed as
\begin{equation}
B = \left(\frac{L_X}{\eta_X c \beta_{\rm rec} R^2}\right)^{1/2} \!\!\!\!\! \simeq 1.3 \, L_{X,43}^{1/2} \eta_{X,-0.3}^{-1/2} \beta_{\rm rec, -1}^{-1/2} R_{13.8}^{-1}~\rm kG.
\end{equation}
Here we normalize the radius $R$ in units of $R_{13.8}=R/1.5 r_g$. The magnetic energy density is then related to the X-ray photon energy density as $u_B = u_X/(6 \beta_{\rm rec} \eta_{X}) \approx 3 u_X \beta^{-1}_{\rm rec, -1} \eta^{-1}_{X,-0.3}$. The corresponding magnetization parameter for the lepton pairs $\sigma_e=B^2/(4\pi n_e m_e c^2)$ is in the range $10-100$ for the inferred values of $L_X$ and $n_e$ (assuming a corona with $\tau_T = 0.5$).
% This is slightly below the upper limits around $\sigma_e\sim 10^4$~\citep{Kimura:2022try}, and consistent with the typical values adopted, e.g., in~\citep{Groselj:2023bgy}.

The power in relativistic protons injected into the corona is also a fraction of the Poynting luminosity, and can be written as
\begin{equation}
L_p = \frac{\eta_p}{\eta_X}  L_X \simeq L_X \frac{\eta_{p,-0.3}}{\eta_{X,-0.3}}.
\end{equation}
The differential energy distribution of relativistic protons is modeled by a broken power law, as described in Sec.~\ref{sec:coronal_model},
\begin{equation}
    \frac{{\rm d}n_p}{{\rm d}E_p {\rm d}t}=
    q_0\mathrm{min}\left[\left(\frac{E_p}{E_{p,\rm br}}\right)^{-1},\left(\frac{E_p}{E_{p,\rm br}}\right)^{-s}\right],
\end{equation}
where $q_0$ is a normalization constant that is determined by equating $\int {\rm d}E_p  E_p \frac{{\rm d}n_p}{{\rm d}E_p {\rm d}t}$ to $L_p$. The power law extends to a maximal energy $E_{p,\rm rad}$, which is determined by the balance of the proton acceleration and cooling rates (see Fig.~\ref{fig:timescales}).
The acceleration timescale that we introduced in Sec.~\ref{sec:coronal_model} can be expressed in terms of $L_X$ as
\begin{eqnarray} 
    t_{\rm acc}\simeq 8.5\times 10^{-7}\;\frac{E_p}{1\;\mathrm{GeV}}\; \left(\frac{\eta_{X, -0.3}}{\beta_{\rm rec,-1} L_{X,43}}\right)^{1/2} R_{13.8} \; \mathrm{s}.
    \label{eq:tacc}
\end{eqnarray}
The energy loss timescales due to proton synchrotron radiation and photohadronic interactions are also related to the X-ray corona emission \citep[for details see][]{
Karavola:2024uui}.

\begin{table} 
    \centering
    \renewcommand{\arraystretch}{1.2}
    \begin{tabular}{lccc}
        \toprule
        Parameter & TXS (LL) & TXS (HL) & TXS (EL) \\
        \midrule
        $L_X$ [$10^{43}$ erg/s] & 4 & 40 & $2\times 10^3$ \\
        $L_p$ [$10^{43}$ erg/s] & 0.8  & 80 & $4\times 10^3$ \\
        $B$ [kG] & 2.6 & 8.1 & 57.3\\
        $\eta_p$ & 0.1 & 1 & 1\\
        \bottomrule
    \end{tabular}
    \caption{Astrophysical parameters for \txs, in the three scenarios discussed in the text. For all cases, we use a coronal size $R\simeq 1.4 r_g=6.4\times 10^{13}$~cm, corresponding to $L=5r_g$.
    }
    \label{tab:astrophysical_params}
\end{table}

Using these relations, we obtain the values summarized in Table~\ref{tab:astrophysical_params} for the main parameters used in the simulations of the radiative emission from the corona. We consider two scenarios that are meant to bracket a conservative and an optimistic scenario for multi-messenger steady coronal production: a low-luminosity (LL) case, in which we assume the lowest X-ray luminosity inferred in Sec.~\ref{sec:observational_X_ray} and a proton luminosity fraction $\eta_p=0.1$, as extracted from PIC simulations; and a high-luminosity (HL) case, in which we assume the highest X-ray luminosity inferred in Sec.~\ref{sec:observational_X_ray}, as well as $\eta_p=1$, ten times higher than what is suggested from PIC simulations and already in tension with the constraints from the corona energetics, which would require $\eta_p+\eta_X<1$. 

In addition, we consider a third scenario, which is meant to test specifically the hypothesis that the 2014/15 neutrino flare might have a coronal origin. In order to do this, we assume that the corona might have been flaring with an X-ray luminosity much higher than the fiducial values inferred in Sec.~\ref{sec:observational_X_ray}. Therefore, to obtain the most optimistic scenario for coronal neutrino production, we choose $L_X$ so as to saturate the 2014/15 upper limits on the X-ray from \textit{MAXI} at a photon energy of $10$~keV~\citep{Petropoulou:2019zqp}, and in addition we choose $\eta_p=1$. For this reason, we refer to this case as an extreme luminosity (EL) scenario.

In all three cases, for the proton injection spectrum we consider a \textit{high-guide-field} (HGF) case, in which $s=3$, and we choose $\sigma_p= 10^7$ ($E_{p,\rm br}\simeq 9.4$~PeV). This ensures that neutrinos can be produced up to PeV energies, as observed by IceCube. Neutrino production at PeV can also be obtained by assuming a \textit{low-guide-field} (LGF) scenario, in which $s=2$, and we choose $\sigma_p=10^5$; results for this alternative case are very similar to the HGF one and we show them in Appendix~\ref{app:lgl}. These two cases, which are consistent with a pair-dominated corona ($\sigma_p \gg \sigma_e$), are meant to show the impact of $\sigma_p$, one of the main unconstrained parameters of our model; see~\citet{Karavola:2024uui} for a systematic discussion on the effects of $\sigma_p$ on the neutrino spectra.

\begin{figure*}
    \includegraphics[width=\textwidth]{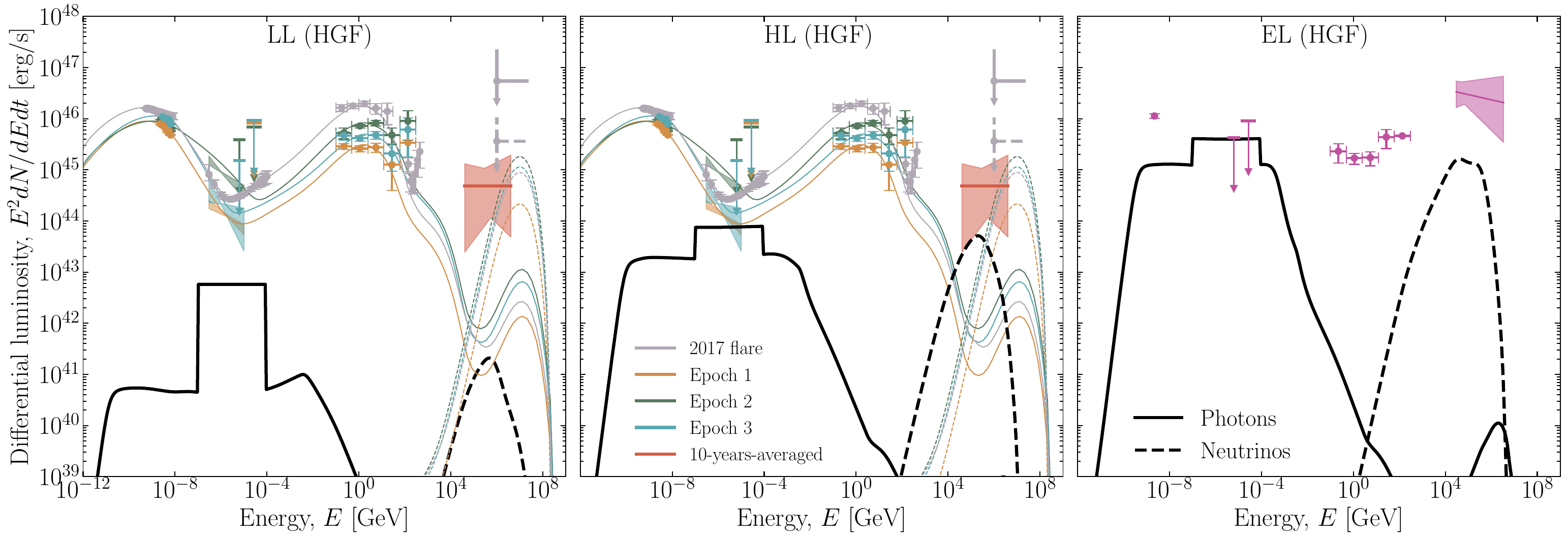}
    \caption{Multi-messenger emission from \txs. We show the coronal emission with solid thick lines, and the blazar emission in different epochs from~\cite{Petropoulou:2019zqp} (for epoch 1, the model with $L_p=L_p^{\rm (max)}$; for epoch 2, the model with the higher X-ray flux) and for the 2017 flare from~\cite{Keivani:2018rnh} (model LMBB2b) as thin lines. For each epoch, we report the measurements across the electromagnetic spectrum~\citep[][and references therein]{Petropoulou:2019zqp,Keivani:2018rnh}; the upper limits on the neutrino flux for the single neutrino event in the 2017 flare~\citep{IC_TXS_2017}, the inferred 10-years-averaged neutrino flux~\citep{IceCube-NGC1068}, and the inferred neutrino flux to explain the 2014/15 flare~\citep{IC_TXS_2014}. Neutrino fluxes are all-flavors; the measured SEDs have been converted to isotropic luminosities using the luminosity distance $d_L\simeq1774$~Mpc, inferred from the redshift of \txs \ \citep[$z=0.3365\pm0.0001$,][]{Paiano_2018}.}\label{fig:sed}
\end{figure*}

In Fig.~\ref{fig:sed} we show the neutrino and electromagnetic emission from the putative corona of \txs \ for the three scenarios.
For comparison, we also show the measurement of the blazar spectral energy distribution (SED) over multiple epochs of emission, extracted from~\cite{Petropoulou:2019zqp}, and during the 2017 flare~\citep{Keivani:2018rnh}.In this way, we can directly understand how the coronal emission compares to the blazar one both during flaring and non-flaring states. We also show in red the neutrino flux reported by IceCube as inferred from the 10-year neutrino point source analysis~\citep{IceCube-NGC1068}. The neutrino flux required to explain the single neutrino event observed in 2017 depends on the assumed time frame for neutrino emission. We show the upper limits displayed in Fig.~4 of~\cite{IC_TXS_2017} assuming $0.5$~yr and $7.5$~yr exposure. Finally, in the third panel, we show the available observations for the 2014/15 epoch, including the neutrino spectrum inferred by IceCube for this period.

In the LL scenario, the neutrino signal from the corona is far below the 2017 flux upper limit, and the neutrino signal averaged over ten years. The SED is composed of the X-ray corona spectrum, injected between $100$~eV and $100$~keV, and of the cascade spectrum, which has a cutoff at about $1$~MeV due to pair annihilation. Across the entire electromagnetic spectrum, the SED of the corona is far below the jet's emission.

In the HL scenario, which is the most optimistic one for steady neutrino production -- and in fact assumes $\eta_p=1$ which is already disfavored by PIC simulations -- we find that the neutrino signal still cannot reach the 10-year average IceCube flux. However, in this case, the coronal X-ray emission is comparable with the jet emission in epochs of low jet activity (epoch~1 and~3). This offers an additional observational opportunity to differentiate between a blazar and a coronal origin for the IceCube neutrinos, since the predicted X-ray emission from the two regions has very different spectrum.

Finally, in the EL case, despite extreme assumptions, both in X-ray luminosity and in the choice $\eta_p=1$, the neutrino flux is unable to match the signal reported by IceCube in 2014/15. Since the X-ray luminosity is already saturating the available upper limits, clearly a further increase either of the X-ray or the proton luminosity would produce a strong tension with such upper bounds. We conclude that a coronal emission for the 2014/15 flare is also challenging to reconcile with the observed neutrino flux.

\section{Discussion}\label{sec:discussion}

The origin of neutrino emission from \txs \ is a key question for neutrino astrophysics. So far, it has been interpreted as a clear hint that blazar jets are high-energy neutrino emitters. Therefore, a scenario in which the neutrino signal originates from the core region of \txs \ instead, similar to NGC~1068, would have fundamental impact on neutrino astrophysics as a whole. In this work, we examine the viability of such a scenario.

We focus on a recently proposed scenario in which protons are accelerated in a reconnection layer formed in the magnetospheric region of the central black hole. This seems to be the most optimistic choice to explain the high-energy neutrinos of \txs, as it provides a fast acceleration mechanism. Other acceleration processes are much slower and the higher energy protons are more strongly affected by cooling. For example, in stochastic acceleration in magnetized turbulence, the typical timescale for proton energization is $t_{\rm acc}\simeq R/c$~\citep{Fiorillo:2024akm}, which is much longer than the acceleration due to the reconnection electric field in the magnetospheric layer (see Eq.~\ref{eq:tacc}). Even considering a wider corona with $R=50r_g$, where the target photon field is more diluted, we have explicitly verified that p$\gamma$ cooling impedes proton acceleration above about 1 PeV (see Appendix~\ref{sec:stochastic}). 

The reconnection scenario, despite being able to accelerate protons to energies above tens of PeV, still faces significant challenges from energetic arguments. Previous works on the subject~\citep{2024arXiv241114598K, Yang:2024bsf} have assumed an X-ray luminosity as large as the observed variable X-ray emission (see e.g. X-ray measurements in Fig.~\ref{fig:sed}), but the latter should truly be regarded as an upper limit on the coronal X-ray luminosity, since observations are plausibly dominated by the jet contribution. Here, we consider the masquerading nature of \txs \ and use its estimated disk luminosity to obtain the coronal X-ray luminosity, using known empirical correlations that are obtained from large samples of bright quasars. On the basis of these correlations, the X-ray luminosity is expected to be one to two orders of magnitude lower than the jet X-ray luminosity. 

Our results show that the steady neutrino emission from \txs, as well as the neutrino flux inferred from the 2017 flare, are challenging to reconcile with coronal emission, both in the conservative LL scenario and in the optimistic HL one. In fact, in the HL scenario, despite the neutrino emission being unable to reach the inferred IceCube flux, the coronal X-rays are competing with the blazar emission in non-flaring epochs. This suggests that multi-component tests of the X-ray emission might provide further constraints on the total neutrino emission from the corona.

We also consider the possibility that in 2014/15 the coronal X-ray luminosity might have been much higher, due to flaring activity, to test whether the neutrino excess reported by IceCube in this period might have a coronal origin.  Similar scenarios, in which protons in the blazar jet interact with dense X-ray external photon fields have also been discussed in the context of the 2014/15 neutrino excess of \txs, \citep{Reimer2019, Petropoulou:2019zqp} and other likely associations of neutrino alerts with blazars \citep{Petropoulou_2020_extreme}. However, even considering an extreme X-ray luminosity which saturates the available upper bounds, and an extreme proton luminosity already disfavored by PIC simulations, the neutrino flux cannot match the reported excess. Furthermore, for such an extreme X-ray luminosity, one would have expected a significantly brighter optical-ultraviolet luminosity as well, potentially leading to tensions with the available measurements in this lower frequency range. We do not explore this question in greater detail, since even such an extreme assumption is unable to match the IceCube neutrino signal. Note that the description of the 2014/15 flare is challenging for the blazar scenario as well~\citep{Murase2018, Reimer2019, Rodrigues2019, Petropoulou:2019zqp}.

The discrepancy with previous works~\citep{2024arXiv241114598K, Yang:2024bsf} primarily comes from the stringent observational constraints on the coronal X-ray luminosity. While \cite{2024arXiv241114598K}~explicitly addresses the corona as neutrino production site, \cite{Yang:2024bsf} refers instead to the accretion flow in the inner regions close to the supermassive black hole. However, the qualitative features of the model are essentially very similar. \cite{Yang:2024bsf} assumes an acceleration timescale comparable with the gyroradius of the proton (divided by $c$); while this is not associated with a specific acceleration mechanism, such a rapid energization cannot happen via stochastic acceleration in turbulence, and must instead be powered by magnetic reconnection. Therefore, our generic constraints on the X-ray and the proton luminosity apply. In particular, the proton luminosity assumed in~\cite{Yang:2024bsf}, of the order of $L_p\sim 10^{46}$~erg/s, is about two orders of magnitude higher than $L_p$ in our HL scenario, and is therefore difficult to reconcile with the expected near-equipartition between X-rays and protons within a reconnection layer. In addition, if the accretion flow is responsible for neutrino production, then super-Eddington accretion may raise additional tension with acceleration of protons at large energies, since the increased density may turn the plasma collisional and hinder efficient acceleration. However, it is difficult to substantiate these statements without a concrete estimate of the acceleration rate, which depends on the assumed mechanism. Our arguments, based on energetics, are independent of these choices.

The jet and coronal emissions may exhibit different time variabilities. Coronal emission, driven by a reconnection layer, is expected to exhibit intermittency on timescales of tens of minutes to a few hours, as the layer is destroyed and reformed over a free-fall time. In contrast, the blazar emission shows variability due to time-variable energy dissipation in the jet, which might be linked to a similar dissipation in the corona. Without a model for the energy dissipation mechanism, we cannot directly relate the time structures of the two signals, making energetics the most reliable argument against coronal emission.

Our arguments are based on the single blazar \txs, but their implications extend to the relative contribution of blazars and non-jetted AGN to the astrophysical neutrino flux. A complementary approach could involve a population-wide study of the expected number of high-energy neutrino associations with blazars and non-jetted AGN. This would rely on luminosity functions, redshift evolution, and assumptions about the populations, making it sensitive to specific production scenarios and uncertainties in population properties. Therefore, we leave this question for future work.

In general, our results reinforce the hypothesis that the neutrino emission of \txs \ has jet rather than coronal origin. In the case of the 2014/15 flare, neither the jet nor the coronal model is able to explain the large expected neutrino signal consistently with the stringent X-ray upper bounds. Although we performed a detailed analysis of the radiative emission, we emphasize that the arguments for this conclusion are at their core very simple, relying on the unavailability of an energy budget in the corona sufficient to explain the large neutrino flux and neutrino energies observed from \txs.

\section{Summary and conclusions}\label{sec:summary}

In light of recent observations of neutrinos from AGN cores, such as NGC 1068, we have studied if the neutrino emission from the AGN blazar TXS 0506+056 may also come from the core only - focusing on corona models. We have demonstrated that stochastic acceleration renders insufficient proton energies, while magnetic reconnection can be used as a plausible acceleration scenario in the corona. Based on the hypothesis that TXS 0506+056 is a masquerading BL Lac, we have carefully assessed the coronal X-ray luminosity, and identified plausible benchmark scenarios. 

Our results clearly demonstrate that the coronal neutrino emission from TXS~0506+056 cannot produce a neutrino flux high enough to produce the observed neutrino emission, neither for the 2017 gamma-ray-flare associated event, nor the 2014-15 neutrino flare. A direct comparison with AGN blazar models, where the model parameters are strongly constrained by the multi-wavelength spectrum, illustrates that the neutrino emission from the jet can be plausibly higher.

While our result provides strong indication that the TXS~0506+056 neutrinos do not come from the corona, they are also no direct evidence for emission from the jet. However, given the history of the discovery associated with a gamma-ray flare, the AGN blazar hypothesis seems to persist as the most plausible scenario. Our finding has profound consequences: presenting one strong counter-example rules out the hypothesis that all observed AGN neutrinos must come from the corona -- and that the jet paradigm is still alive! 

\section*{Acknowledgements} 
We thank Margot Boughelilba for useful comments on this manuscript. The research project has been funded by the ``Program for the Promotion of Exchanges and Scientific Collaboration between Greece and Germany IKYDA--DAAD'' 2024 (IKY project ID 309; DAAD project ID: 57729829). 
D.F.G.F. is supported by the Alexander von Humboldt Foundation (Germany).  M.P. acknowledges support from the Hellenic Foundation for Research and Innovation (H.F.R.I.) under the ``2nd call for H.F.R.I. Research Projects to support Faculty members and Researchers" through the project UNTRAPHOB (Project ID 3013).

\appendix
\section{Stochastic acceleration driven by turbulence}\label{sec:stochastic}

\begin{figure*}
    \begin{center}        
    \includegraphics[width=0.9\textwidth]{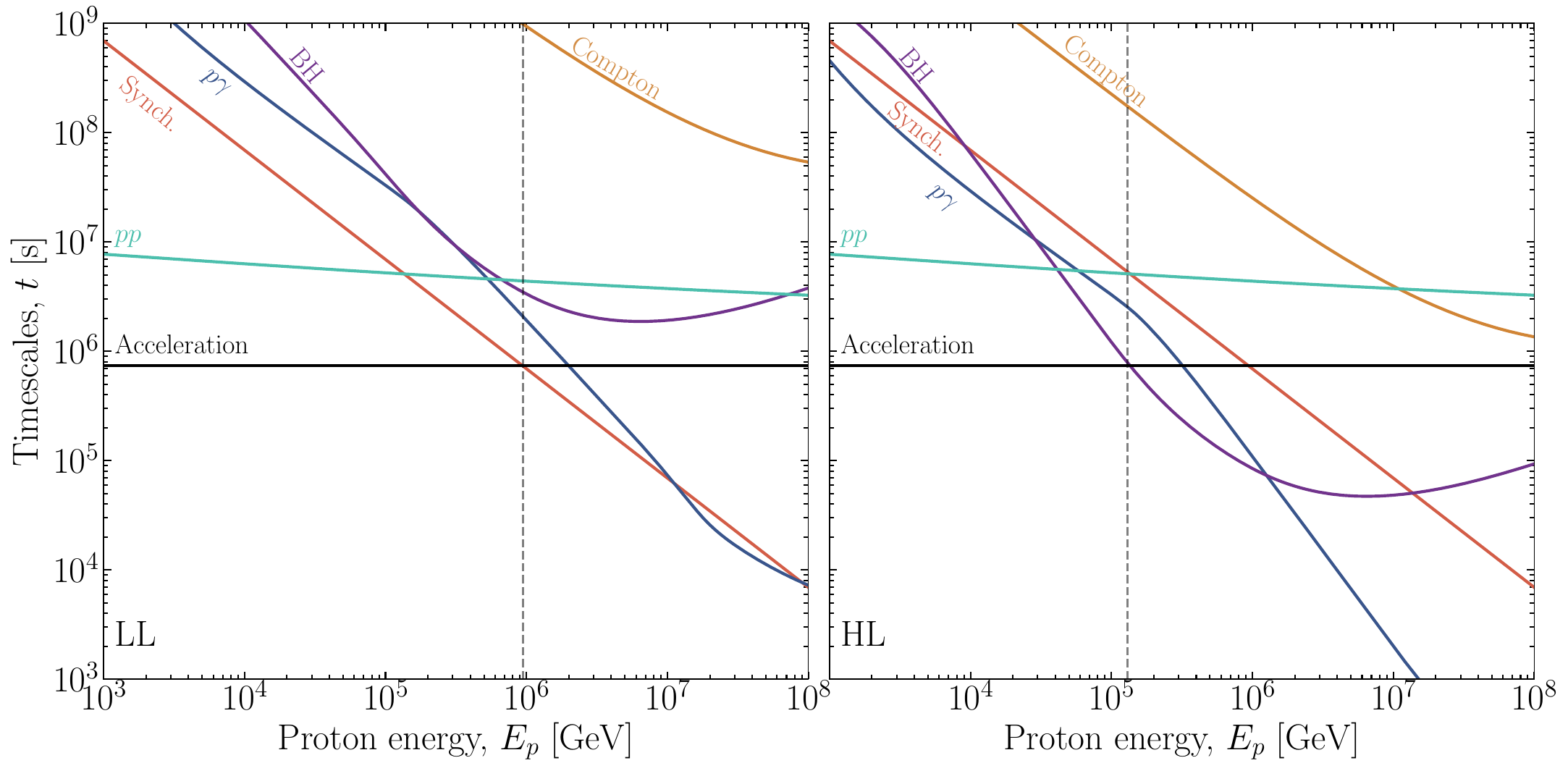}
    \end{center}
    \caption{Acceleration and cooling timescales for the scenario with stochastic acceleration; we show separately the LL (left panel) and HL (right panel) case identified in the text. The vertical dashed lines identify the maximal energies that can be achieved in each case. The coronal size is fixed to $R=50r_g$.}\label{fig:stochastic}
\end{figure*}

If the corona is relatively large ($R\sim 10-100$~$r_g$), we can envision it as a more or less spherical environment permeated with X-rays and optical-ultraviolet (OUV) photons from the accretion disk. Because of the relatively large radius, the typical compactness of the corona for MeV photons is not necessarily very large, which suggests that the corona may not be strongly pair-dominated. Thus, for simplicity, we may assume that the number density of protons $n_p$ is equal to the lepton number density $n_e$; since the Thomson opacity of the corona is $\tau_T\simeq 0.5$ we have
\begin{equation}\label{eq:lepton_number_density}
    n_p=n_e\simeq 1.7\times 10^{10}\;\frac{r_g}{R}\;\mathrm{cm}^{-3}.
\end{equation}

As in~\cite{Fiorillo:2024akm}, we assume a magntization $\sigma$, defined as the ratio between the magnetic and the rest-mass energy density, of order unity
\begin{equation}
    \sigma=\frac{B^2}{4\pi(n_e m_e+n_p m_p)c^2},
\end{equation}
which leads to an estimated magnetic field
\begin{equation}
    B\simeq 1.8\times 10^4\;\left(\frac{\sigma r_g}{R}\right)^{1/2}\;\mathrm{G}.
\end{equation}
We should generally distinguish between the total magnetic field and its turbulent component, but we are mainly interested in order-of-magnitude estimates, which suffice to show the challenges that a stochastic model would encounter in explaining the neutrino observations. Therefore, we neglect this distinction in what follows.
If acceleration proceeds via stochastic energization in the stationary turbulence within the corona, then its typical timescale~\citep{CS19,Fiorillo:2024akm} is $t_{\rm acc}=10 \ell/c \sigma$, where $\ell$ is the coherence length of the turbulence, which in order of magnitude we may take to be comparable with the radius $\ell \sim R$. Notice that this acceleration timescale is much longer than the one considered in~\cite{Yang:2024bsf}, which is more representative of the reconnection scenario we discuss extensively in the main text. Thus, within the corona, it is extremely challenging to find a scenario in which protons can be consistently accelerated up to energies of the order of $10$~PeV, which would be needed to explain the neutrinos observed by IceCube in the $100$~TeV-$1$~PeV range. 

To assess this more conclusively, we obtain the cooling timescale for protons with energy $E_p$ and compare it with the acceleration timescale. The cooling timescale depends very sensitively on the photon field within the corona: here we consider an AGN photon field parameterized as in~\cite{Marconi:2003tg}, which depends on a single parameter, which can be taken to be the X-ray luminosity $L_X$ within the $2-10$~keV band. We consider both the LL and HL scenario here, using the same values of $L_X$ as in the main text. For both cases, we determine the cooling timescale assuming a benchmark value for $R=50r_g$, and we compare it with the acceleration timescale in Fig.~\ref{fig:stochastic}; the choice of a relatively large corona is made to reduce as much as possible the cooling losses. We obtain the timescales using the numerical code \texttt{AM$^3$}~\citep{Klinger:2023zzv}; we have verified that using the analytical expressions in~\cite{Fiorillo:2024akm} leads to the same results. 

Even in the LL case, where the losses due to proton-photon collisions are as small as possible, proton synchrotron is rapid enough to limit proton acceleration to about $E_p\lesssim 1$~PeV. Thus, neutrinos could not be produced with energies above about $20$~TeV, in clear contradiction with the signal measured by IceCube (see Fig.~5 of~\cite{IceCube:2022der}), which reaches up to a few PeV. For the HL case, the maximal energy is even lower, due to the rapid photohadronic cooling. We conclude that the observed energy range for the IceCube neutrinos is impossible to explain within a model of stochastic acceleration.

\section{Low-guide-field scenario}\label{app:lgl}

\begin{figure*}
    \includegraphics[width=\textwidth]{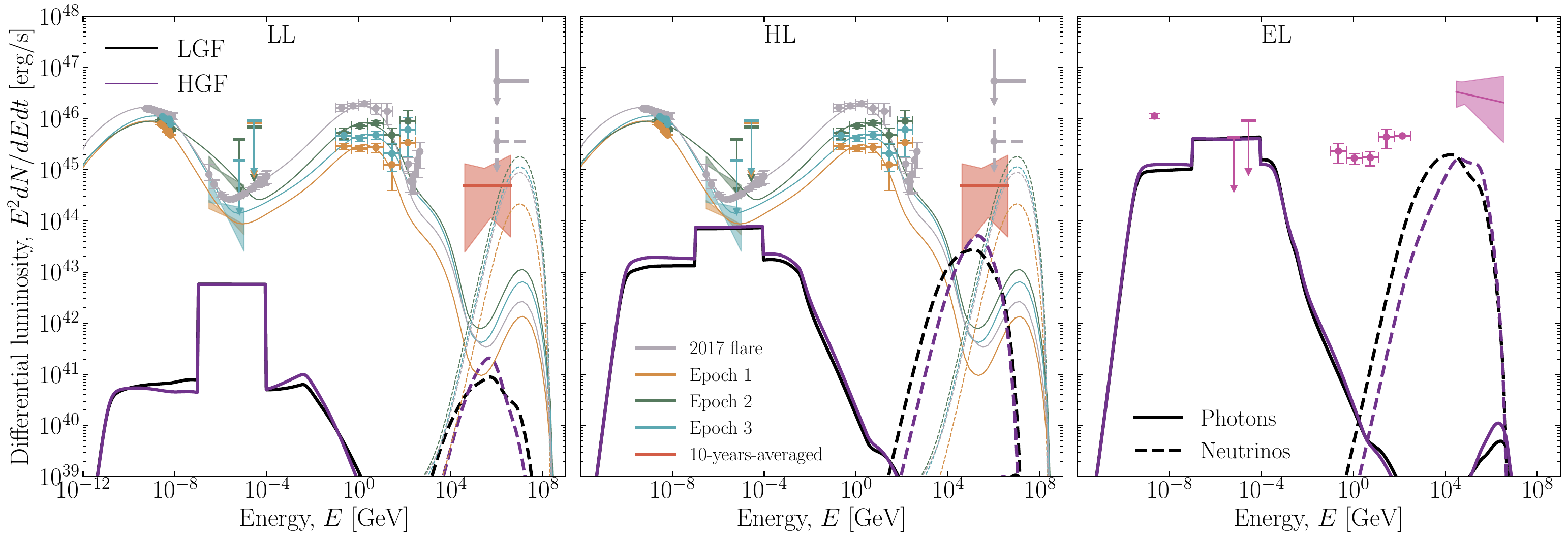}
    \caption{Same as Fig.~\ref{fig:sed} in the main text, showing the results for both the LGF and HGF proton injection spectrum.}\label{fig:sed_lgf}
\end{figure*}

In this section, we show for completeness in Fig.~\ref{fig:sed_lgf} how the results in the main text when an injection scenario with low guide field is considered. In this case, protons are injected with a $\sigma_p=10^5$, corresponding to a break energy of about $100$~TeV, and have a power-law index $s=2$ at higher energies.

The differences in the produced particles is minimal, and is mainly visible in the neutrino spectrum. In the LGF scenario, the photohadronic bump of high-energy neutrinos is broadened, as a consequence of the protons having a flat energy injection spectrum. The form of the neutrino spectrum is also distorted close to the peak, mostly due to the cooling of the secondary pions and muons. However, all of these differences have a minor effect in the development of the radiative cascade, since the amount of energy injected in the form of high-energy photons and pairs is very similar, and their subsequent reprocessing to lower energies is insensitive to the details of the proton injection spectrum.

\section{Components of the radiative cascade}\label{app:cascade}
\begin{figure*}
    \includegraphics[width=\textwidth]{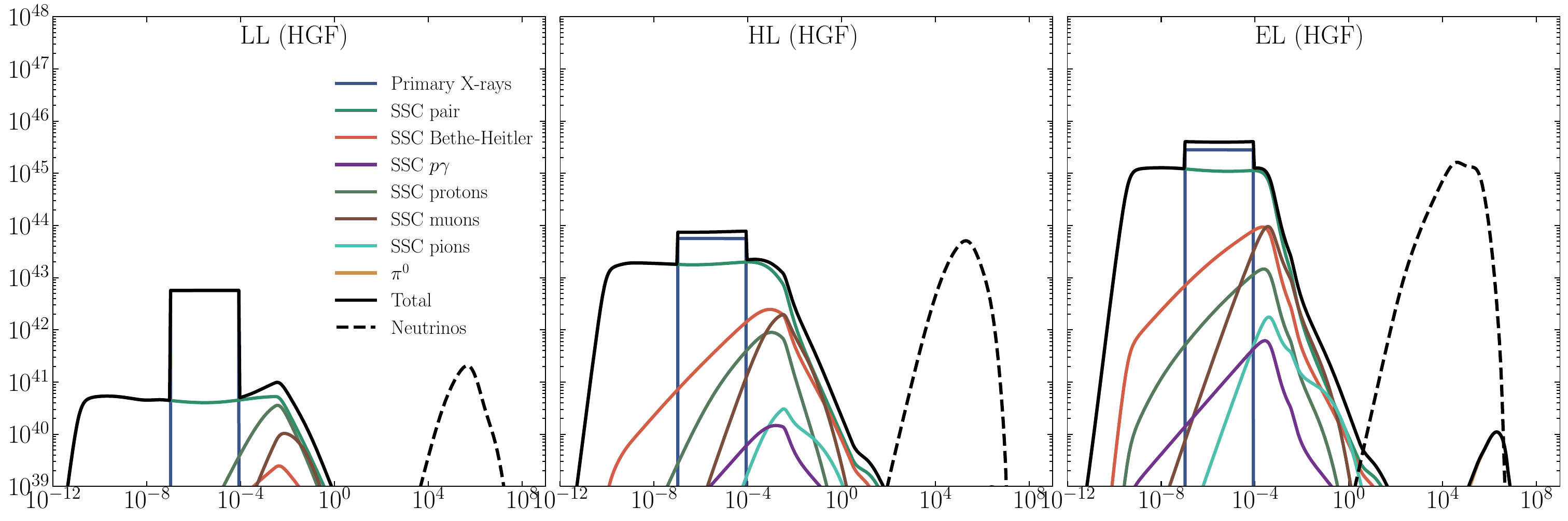}
    \caption{Multi-messenger emission in all three scenarios (LL, HL, EL) considered in the main text. We decompose the emission into its components, including the primary X-rays injected in the simulation and the components of the radiative cascade discussed in the text.}\label{fig:cascade}
\end{figure*}

The simulations of this work are performed using the numerical code \texttt{AM$^3$} \citep{Klinger:2023zzv}, which can separately track the population of particles stemming from each of the physical processes considered. Fig.~\ref{fig:cascade} shows an in-depth overview of the different components of the SEDs obtained in our model, for the HGF case shown in Fig.~\ref{fig:sed}.
Due to the high density of coronal x-rays (labeled as \textit{primary X-rays} in Fig.~\ref{fig:cascade}), the production of electron-positron pairs via $\gamma \gamma$ annihilation is the most efficient process for photons with $E_\gamma \gtrsim \mathrm{MeV}$. Hence, the cascaded spectrum has a sharp cutoff at higher energies, and the synchrotron self-compton (SSC) emission from these secondary pairs (\textit{SSC pair}) dominates the electromagnetic spectrum. Non-thermal protons contribute as well to the photon emission, directly, via their SSC emission (\textit{SSC protons}), and indirectly, via the secondaries from proton-photon scattering. The latter introduces two channels of emission: Bethe-Heitler (BH), which leads to direct production of pairs with resulting SSC emission (\textit{SSC Bethe-Heitler}); and $p\gamma$, which produces pairs, with resulting SSC emission (\textit{SSC $p\gamma$}), charged pions (leading to \textit{SSC pions}) and muons (leading to \textit{SSC muons}), as well as neutral pions promptly decaying to photons (labeled as $\pi^0$).

\bibliographystyle{aasjournal}
\bibliography{References}

\end{document}